\documentclass[aps,prb,reprint,superscriptaddress]{revtex4-2}
\usepackage[T1]{fontenc}
\usepackage[utf8]{inputenc}
\usepackage{graphicx}
\usepackage{booktabs}
\usepackage{array}
\usepackage{amsmath,amssymb}
\usepackage{bm}
\usepackage{longtable}
\usepackage{multirow}
\usepackage{dcolumn}
\usepackage{xcolor}
\usepackage[hidelinks]{hyperref}
\usepackage{titlesec}
\renewcommand\thesection{\Roman{section}}
\titleformat{\section}{\normalfont\large\bfseries}{\thesection.}{0.6em}{}

\begin{document}

\title{Glassy magnetic freezing of interacting clusters in LK-99-family materials}

\author{Serafim Teknowijoyo}
\affiliation{Laboratory of Advanced Quantum Materials, Institute for Quantum Studies, Chapman University, Orange, CA 92866 USA}
\author{Domenico Napoletani}
\affiliation{Germinal Computing LLC, Fairfax, VA 22033, USA}
\author{Vahan Nikoghosyan}
\affiliation{Laboratory of Advanced Quantum Materials, Institute for Quantum Studies, Chapman University, Orange, CA 92866 USA}
\affiliation{Institute for Physics Research, National Academy of Sciences, Ashtarak-2, 0304 Armenia}
\author{Armen Gulian}
\email{gulian@chapman.edu}
\affiliation{Laboratory of Advanced Quantum Materials, Institute for Quantum Studies, Chapman University, Orange, CA 92866 USA}

\begin{abstract}
We report reproducible magnetization anomalies appearing below room temperature in copper-doped apatite materials belonging to the LK-99 family synthesized via hydrothermal methods. These anomalies are observed consistently across samples prepared under comparable conditions. Although the extracted Mydosh parameter lies within the range often associated with vortex-glass behavior in superconductors, a detailed analysis of DC magnetization, AC susceptibility, field dependence, and magnetic memory effects demonstrates that the observed phenomena are not related to superconductivity. Instead, the data are consistent with glassy magnetic freezing of interacting clusters. Compositional and structural analysis identifies covellite (CuS), an ubiquitous secondary phase in these intrinsically multiphase materials, as the primary origin of the observed behavior. Our results clarify the magnetic origin of LK-99-related anomalies and highlight the importance of phase complexity in interpreting apparent superconducting signatures in this materials family.
\end{abstract}

\maketitle

\section{Introduction}
In 2023, Lee, Kim, and Kwon claimed the discovery of a superconductor operating at room temperature and ambient pressure \cite{ref1}. This announcement elicited an intense response from both theoretical and experimental communities. On the theoretical side, it stimulated broad speculation regarding possible mechanisms for superconductivity in the exotic crystal structure of modified apatite \cite{ref2,ref3,ref4,ref5,ref6,ref7,ref8,ref9,ref10,ref11,ref12,ref13,ref14,ref15,ref16,ref17}. Experimentally, however, numerous attempts to reproduce the reported results were unsuccessful \cite{ref18,ref19,ref20,ref21,ref22,ref23,ref24,ref25,ref26,ref27,ref28,ref29,ref30,ref31,ref32}. These references arrived within the first two months after the original preprint \cite{ref1} and later appeared as journal publications.

A central difficulty lay in the fact that the exact chemical composition of the reported material remained unclear, even to the original authors. Moreover, the synthesis reaction described in Ref.~\onlinecite{ref1} was chemically unbalanced, leaving several key questions unresolved. Subsequent efforts by many independent groups expanded the characterization of this system \cite{ref33,ref34,ref35,ref36,ref37,ref38,ref39,ref40,ref41,ref42,ref43,ref44,ref45,ref46,ref47,ref48,ref49,ref50,ref51,ref52,ref53,ref54,ref55} and ultimately led to the conclusion that the reported ``superconducting'' signatures can be explained by a structural phase transition in Cu$_2$S \cite{ref28}, one of the constituent phases in this inherently multiphase material. This interpretation largely resolved the initial controversy surrounding the reported superconducting signatures and effectively concluded the disputes.

Depending on the chosen synthesis route---and, critically, on subtle variations in synthesis parameters (for example, the pH during hydrothermal growth)---the resulting materials exhibit not only quantitative but also qualitative differences. These differences extend beyond chemical composition to include crystal structure as well as magnetic and transport properties. This diversity motivates the conditional grouping of all such multiphase materials produced via different modified-apatite synthesis protocols under the collective designation ``LK-99.'' Throughout this work, we use the term ``LK-99 family'' to denote chemically and structurally distinct, yet synthesis-related, multiphase materials rather than a single well-defined compound.

\begin{figure}[t]
\includegraphics[width=0.75\columnwidth]{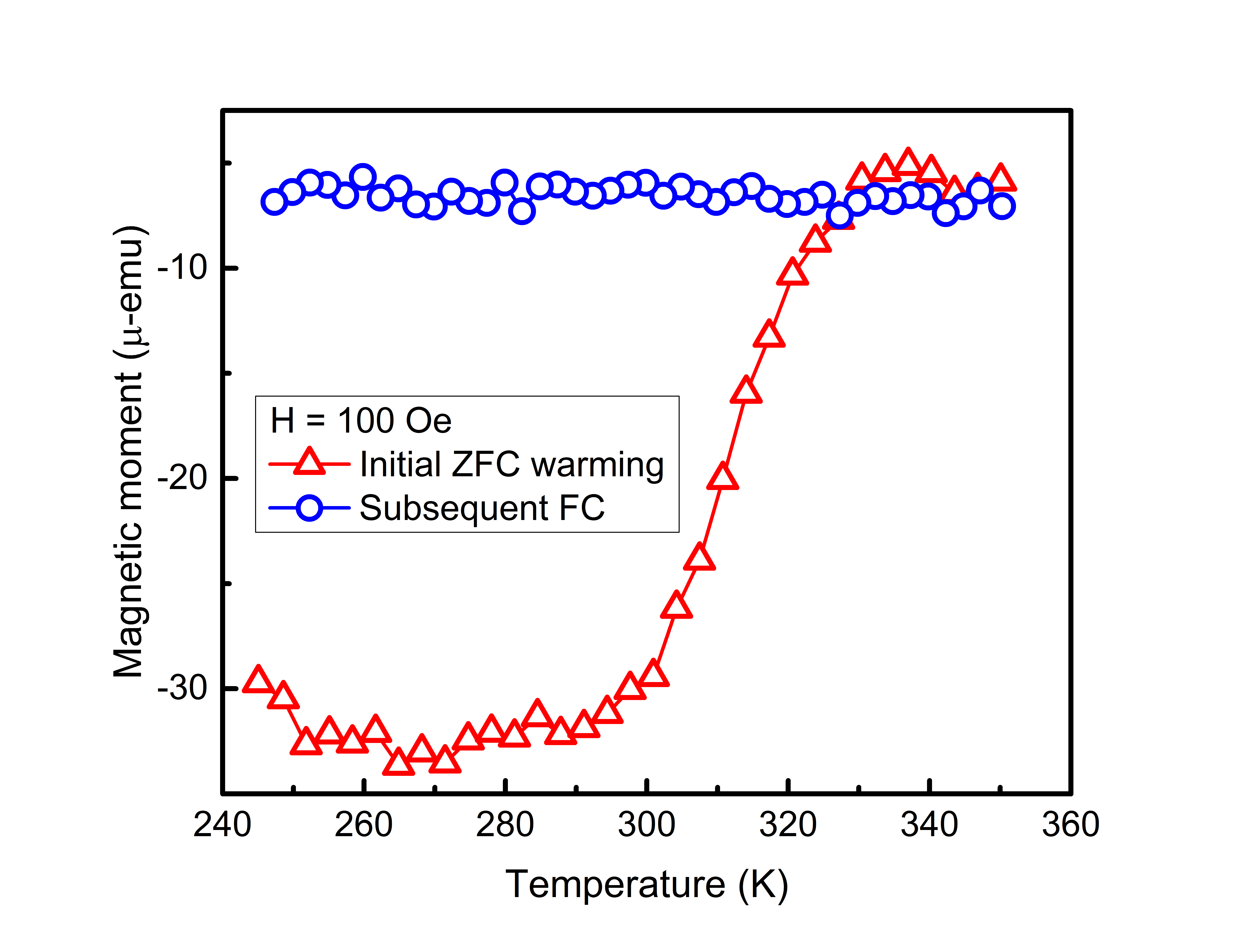}
\caption{Temperature dependence of the magnetization measured in a copper-doped pyromorphite pellet (mass $\approx 98.5$ mg, diameter $\approx 4$ mm, thickness $\approx 1.5$ mm). Zero-field-cooled (ZFC) warming and subsequent field-cooled (FC) data are shown, illustrating an anomaly near room temperature. All other measured pellets have the same diameter.}
\label{fig:fig1}
\end{figure}

In the present work, we set aside the question of superconductivity at or above 300 K, despite a very intriguing behavior above room temperature in some of our samples (as shown in Fig.~\ref{fig:fig1}); it is out of the scope of this discussion since we admit that extraordinary claims require extraordinary evidence, including reproducibility, long-term stability, and a comprehensive and internally consistent set of physical characteristics, criteria that have not yet been satisfied.

While agreeing that phase transitions in Cu$_2$S can adequately account for the experimental observations reported in Ref.~\onlinecite{ref1}, this conclusion naturally raises a further question: \emph{could materials belonging to the LK-99 family exhibit superconductivity at lower temperatures?}

Motivated by this question, we investigated a series of samples that reproducibly display the anomalous behavior shown in Fig.~\ref{fig:fig2}. Although this response at first glance resembles a Meissner-like effect, a detailed analysis demonstrates that it originates from the melting of a spin-glass state. Furthermore, we obtained experimental evidence indicating that the presence of the CuS phase within LK-99-type materials is responsible for this behavior.

\section{Experimental details and results}
Modified apatite compositions were synthesized using a standard high-pressure hydrothermal method in stainless-steel autoclaves equipped with PTFE-lined inserts of 30 and 100 mL volume. The starting reagents were copper nitrate [Cu(NO$_3$)$_2$], lead nitrate [Pb(NO$_3$)$_2$], ammonium dihydrogen phosphate (NH$_4$H$_2$PO$_4$), and potassium sulfide (K$_2$S). The synthesis protocol follows Ref.~\onlinecite{ref54} and consists of two successive stages.

At the first stage, the reagents were mixed in a molar ratio Cu$^{2+}$ : Pb$^{2+}$ : PO$_4^{3-}$ : S$^{2-}$ = 9 : 1 : 6 : 2. Ammonium hydroxide was added to deionized water to adjust the pH to 8. The resulting solution was maintained in a water bath at 60$^{\circ}$C for 24 h. Subsequently, the mixture was transferred into the hydrothermal reactor, typically filling $\sim$75\% of the reactor volume; in some runs, the filling fraction was increased to as much as 90\% in order to enhance the autogenous pressure during hydrothermal growth. Hydrothermal treatment was then typically carried out at 150--160$^{\circ}$C for 48 h. During this process, the initially brown solution transformed into a gray precipitate suspended in a nearly colorless supernatant. The precipitate was separated by filtration and dried. In addition to rinsing with deionized water, ethyl alcohol was used to accelerate the drying process. The dried material obtained after the first stage was then mixed with an additional amount of sulfide ions (S$^{2-}$) at a molar ratio of 1:10. The pH was again adjusted to 8 using ammonium hydroxide. This mixture was magnetically stirred at 60$^{\circ}$C for 12 h, followed by a second hydrothermal treatment at 150$^{\circ}$C for 8 h or longer. When these conditions were strictly maintained, the resulting precipitate was black, corresponding to the desired modified apatite phase. A summary of the synthesis temperature regimes is provided in Table~\ref{tab:table1}.

\begin{table}[t]
\caption{Regimes of hydrothermal synthesis of 4 samples (a,b,c in 100 mL and d in 30 mL reactor).}
\label{tab:table1}
\begin{ruledtabular}
\begin{tabular}{ccccc}
\# & 1st stage & 2nd stage & Powder & Pellet \\
\hline
 a & 150$^{\circ}$C/48h & 150$^{\circ}$C/11h & $\sim$403 mg & 84.5 mg \\
 b & 160$^{\circ}$C/48h & 160$^{\circ}$C/16h & $\sim$532 mg & 82.5 mg \\
 c & 150$^{\circ}$C/53h & 150$^{\circ}$C/14h & $\sim$777 mg & 58.5 mg \\
 d & 150$^{\circ}$C/48h & 150$^{\circ}$C/24h & $\sim$41 mg  & 40.8 mg \\
\end{tabular}
\end{ruledtabular}
\end{table}

\begin{figure}[t]
\includegraphics[width=0.75\columnwidth]{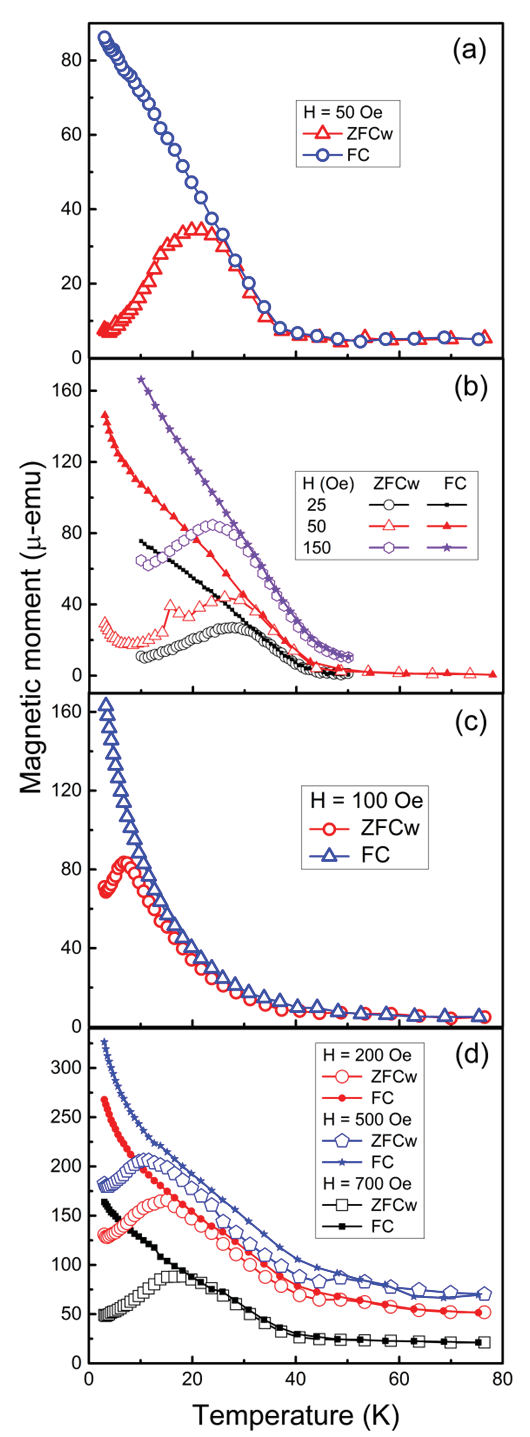}
\caption{Temperature dependence of the magnetization showing bifurcation between zero-field-cooled (ZFC) and field-cooled (FC) curves for four representative samples synthesized under slightly different hydrothermal conditions. Panels (a)--(d) correspond to samples \#a--\#d listed in Table~\ref{tab:table1}. Applied magnetic fields are indicated in each panel.}
\label{fig:fig2}
\end{figure}

The powders were subsequently pressed into pellets, which were used for magnetic characterization. Temperature-dependent magnetization measurements in various applied magnetic fields, as well as AC magnetic susceptibility measurements over a range of frequencies, were performed using a DynaCool physical property measurement system (PPMS, Quantum Design). Panels (a)--(d) in Fig.~\ref{fig:fig2} present the key experimental observation: the bifurcation of zero-field-cooled and field-cooled magnetization curves.

\section{Discussion and conclusions}
These reproducible observations straightforwardly raise a question: may this be a superconducting transition in a multiphase sample superimposed with a parasitic magnetic phase? Is this related with the frozen vortex effect \cite{ref54} or some other effect?

\begin{figure}[t]
\includegraphics[width=0.75\columnwidth]{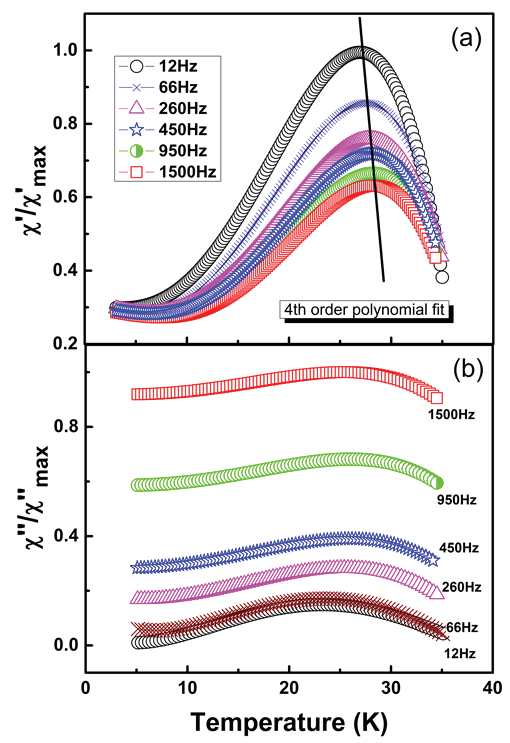}
\caption{Temperature dependence of the real ($\chi'$) and imaginary ($\chi''$) components of the AC magnetic susceptibility for the sample shown in Fig.~\ref{fig:fig2}, measured at several excitation frequencies. The vertical line marks the temperature of the $\chi'$ maxima. Solid line is a guide to the eye.}
\label{fig:fig3}
\end{figure}

To understand the situation better \cite{ref56,ref57}, we picked up one of the samples reported in Fig.~\ref{fig:fig2} (the one shown in panel d) and ran AC susceptibility measurements, Fig.~\ref{fig:fig3}. The real part of the AC susceptibility, $\chi'(T)$, measured at zero DC field with an AC excitation amplitude of 5 Oe, displays a broad cusp centered at $T \approx 27$--28 K. The cusp position shifts systematically toward higher temperature with increasing frequency in the range 12--1500 Hz, accompanied by a modest reduction in amplitude (noticeably, $\chi''(T)$ has the same tendency though curves are in reciprocal order). From the frequency dependence of the $\chi'$ peak temperature $T_f$, the relative shift per decade of frequency (Mydosh parameter \cite{ref58,ref59}) is found to be
\begin{equation}
K \equiv \frac{\Delta T_f}{T_f\,\Delta \log_{10} f} \approx 0.02.
\label{eq:K}
\end{equation}

\begin{figure}[t]
\includegraphics[width=0.75\columnwidth]{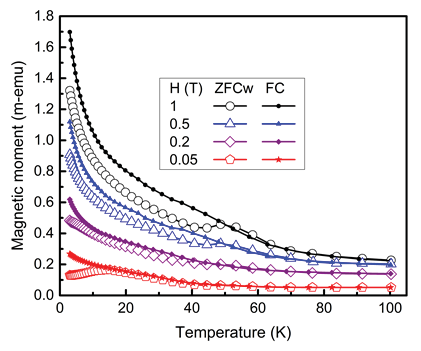}
\caption{Temperature-dependent magnetization of the sample shown in Fig.~\ref{fig:fig2}(d), measured under higher applied magnetic fields (Tesla scale). Zero-field-cooled warming and field-cooled data are shown for different field strengths, illustrating the evolution of the low-temperature anomaly with increasing field.}
\label{fig:fig4}
\end{figure}

This value is substantially larger than that characteristic of canonical metallic spin glasses such as CuMn or CuFe alloys, for which $K\simeq 0.004$--0.006 is well established experimentally \cite{ref59}. For vortex-glass states in disordered superconductors, values in the range $K\approx 0.01$--0.1 have been reported, and that is better matching with Eq.~(\ref{eq:K}). Alternatively, the enhanced value of $K$ observed here can represent cluster-glass or interacting spin-glass systems \cite{ref60}, where magnetic moments freeze collectively in spatially heterogeneous clusters with a broad spectrum of relaxation times. To choose between these alternatives, we performed studies of DC magnetization of this sample in stronger magnetic fields, Fig.~\ref{fig:fig4}. Rather than quenching the effect, at higher values of $H$, the $T_f$ increases, which definitely contradicts the superconductivity scenario.

By contrast, the enhanced value of $K$ observed here is typical of cluster-glass or interacting spin-glass systems. To choose between these two cases, we estimated the characteristic relaxation time extracted from the $\chi'$ peak positions using a critical slowing-down form \cite{ref57,ref61}:
\begin{equation}
\tau = \tau_0\left(\frac{T_f}{T_g}-1\right)^{-z\nu}.
\label{eq:tau}
\end{equation}

Fitting the frequency-dependent peak positions to a critical slowing-down form yields a glass temperature $T_g \approx 24$ K, a dynamic exponent $z\nu \approx 12$, and a microscopic attempt time $\tau_0 \approx 10^{-13}$ s. These parameters are physically reasonable and closely match values reported for insulating and granular spin-glass and cluster-glass systems, including canonical CuMn alloys. The coexistence of a canonical-like microscopic attempt time with a non-canonical Mydosh parameter highlights an important distinction: while the freezing dynamics are collective and glassy in nature, the magnetic entities involved are not individual spins uniformly coupled via long-range interactions, as in CuMn or CuFe alloys, but rather interacting clusters arising from structural, chemical, or electronic inhomogeneity. Such behavior is well documented in disordered Cu-based oxides, doped low-dimensional Cu systems, and granular magnetic materials, where deviations from canonical scaling coexist with otherwise spin-glass-like critical dynamics.

\begin{figure}[t]
\includegraphics[width=0.75\columnwidth]{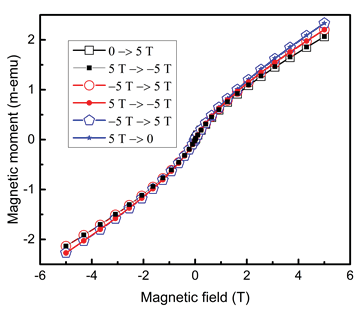}
\caption{Magnetization as a function of magnetic field measured during successive field sweeps between $\pm 5$ T, demonstrating magnetic memory and partial irreversibility. The absence of sharp hysteresis loops and the dependence on field history are characteristic of glassy magnetic behavior.}
\label{fig:fig5}
\end{figure}

To confirm further this conclusion about cluster glass, we analyzed reversibility in strong magnetic fields and discovered the magnetic memory effect, Fig.~\ref{fig:fig5}. Within a cluster-glass framework, each magnetic cluster behaves as an effective superspin interacting weakly but non-negligibly with its neighbors. These interactions generate a complex free-energy landscape containing multiple metastable states. Once the field is swept to $\pm 5$ T, it partially aligns the clusters. Some clusters become trapped in local minima, and when the field is swept back, the system remembers the previous configuration. This produces history-dependent $M(H)$, lack of full reversibility, with the absence of sharp hysteresis loops, as seen in Fig.~\ref{fig:fig5}.

Thus, the presented analysis provides robust evidence for glassy magnetic freezing involving interacting clusters, rather than canonical metallic spin-glass behavior or superconductivity.

\begin{figure}[t]
\includegraphics[width=0.75\columnwidth]{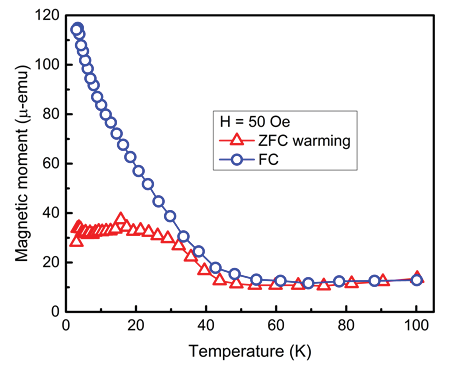}
\caption{Temperature-dependent magnetization of a reference covellite (CuS) pellet (mass $\approx 175$ mg) prepared from commercial powder (Sigma-Aldrich, $\ge 99\%$ metals basis). Zero-field-cooled warming and field-cooled data are shown for an applied field of 50 Oe.}
\label{fig:fig6}
\end{figure}

To figure out the main phase causing the bifurcation, we referred to the compositional and structural data of samples presented in Table~\ref{tab:table2}. As follows from Table~\ref{tab:table2}, all specimens possess a significant amount of covellite. Following this conclusion, we prepared a pellet of CuS and obtained a result shown in Fig.~\ref{fig:fig6}. This was surprising, since we did not find any literature evidence of low-temperature magnetic hysteresis in covellite. Interestingly, there was a claim of superconductivity at 40 K in this system \cite{ref63}, which stimulated a theoretical exploration by Mazin \cite{ref64} on the likelihood of such a phenomenon. This theoretical analysis leaned toward a lack of superconductivity and at the time of its publication the authors of Ref.~\onlinecite{ref65} withdrew their result by finding an alternative explanation of experimental data. We were interested in analyzing this topic using the machine-learning approach to the prediction of physical properties that was introduced in Ref.~\onlinecite{ref66}. To this extent, we explored the Cu-S compositional space in the vicinity of stoichiometric CuS by sampling Cu$_x$S$_y$ compositions in the neighborhood of $(x,y)=(1,1)$. Throughout this region, the calculated stability landscape exhibits a clear and reproducible structure: the most stable compositions cluster along a near-stoichiometric ridge $x\approx y$, i.e., Cu:S$\approx 1{:}1$. This behavior is consistent with the expectation that deviations from charge balance and local coordination environments rapidly increase the decomposition energy (energy above the convex hull) in binary chalcogenides.

In contrast, the distribution of predicted superconducting critical temperatures $T_c$ organizes into distinct off-stoichiometric pockets that are generally displaced from the stability ridge. Consequently, regions of maximal thermodynamic stability and elevated $T_c$ exhibit limited overlap, revealing an intrinsic tradeoff within the Cu$_x$S$_y$ family. Compositions that maximize stability seem to inhibit high-$T_c$, whereas the most promising high-$T_c$ candidates reside in comparatively unstable or metastable regions of the compositional space.

This decoupling between stability and $T_c$ motivates a targeted strategy for materials discovery in the Cu-S system. Rather than restricting the search to the stability ridge near CuS, one can identify high-$T_c$ pockets within the off-stoichiometric Cu$_x$S$_y$ landscape and then seek chemical routes to stabilize these compositions. One practical approach is ternary substitution, which can enhance stability without fully disrupting the electronic structure features responsible for the elevated $T_c$. In particular, substitution with Mo in the Cu-Mo-S system offers a plausible stabilizing mechanism for Cu-rich or S-deficient compositions \cite{ref67}.

\begin{table*}[t]
\caption{Compositional and structural data obtained after second stage of hydrothermal synthesis using electron dispersive spectroscopy (EDS) and x-ray diffraction (XRD). Instruments' details are in Ref.~\onlinecite{ref62}.}
\label{tab:table2}
\begin{ruledtabular}
\begin{tabular}{cll}
\# & Elements (\% at.) & Weight fraction (\%) \\
\hline
 a & Cu: 27.1; S: 3.3; Pb: 1.6; P: 11.5; O: 56.5 & CuS (Covellite): 79.3; PbS (Galena): 3.7; Pb$_{10}$P$_6$O$_{25}$: 17 \\
 b & Cu: 49.9; S: 44.1; Pb: 3.5; O: 2.2; Al: 0.4 & CuS (Covellite): 80.8; PbS (Galena): 19.2 \\
 c & Cu: 44.3; S: 46.8; Pb: 3.9; P: 1.0; O: 3.9 & CuS (Covellite): 80.7; PbS (Galena): 19.3 \\
 d & Cu: 49.6; S: 43.9; Pb: 3.0; O: 3.1; Al: 0.5 & CuS (Covellite): 64.6; PbS (Galena): 35.4 \\
\end{tabular}
\end{ruledtabular}
\end{table*}

In conclusion, we have shown that reproducible low-temperature magnetization anomalies observed in LK-99-family materials arise from glassy magnetic freezing of interacting clusters rather than superconductivity. AC susceptibility, field-dependent magnetization, critical slowing-down analysis, and magnetic memory effects consistently support a cluster-glass interpretation. Structural and compositional analysis identifies covellite (CuS), an unavoidable secondary phase in these multiphase systems, as the dominant contributor to this behavior.

While the LK-99 materials family remains a fertile platform for discovering unusual magnetic and electronic phenomena, our results place strong constraints on interpretations invoking superconductivity. More broadly, the targeted modification of mineral-derived compounds \cite{ref68} via hydrothermal and other methods of synthesis \cite{ref69} remains a promising strategy for exploring complex emergent states of matter.

\begin{acknowledgments}
This research was supported in part by the ONR Grant No.~N00014-24-1-2595. We would like to express our deep gratitude to Tyrel McQueen and Gregory Bassen for their advice with hydrothermal synthesis at the initial stage of this work at Johns Hopkins University. Also, we are grateful to the Physics Art Frontiers and Sergey Abrahamyan for the provided technical assistance.
\end{acknowledgments}


\begin{thebibliography}{69}
\bibitem{ref1} S. Lee, J. H. Kim, Y. W. Kwon, The First Room-Temperature Ambient-Pressure Superconductor, arXiv:2307.12008 (2023).

\bibitem{ref2} Q. Baskaran, Broad Band Mott Localization is all you need for Hot Superconductivity: Atom Mott Insulator Theory for Cu-Pb Apatite, arXiv:2308.01307 (2023).

\bibitem{ref3} W. Chen, Berry curvature and quantum metric in copper-substituted lead phosphate apatite, arXiv:2308.05124 (2023).

\bibitem{ref4} L. Si, M. Wallerberger, A. Smolyanyuk, S. di Cataldo, J. M. Tomczak, and K. Held, Pb$_{10-x}$Cu$_x$(PO$_4$)$_6$O: a Mott or charge transfer insulator in need of further doping for (super)conductivity, arXiv:2308.04427 (2023).

\bibitem{ref5} L. Si and K. Held, Electronic structure of the putative room-temperature superconductor Pb$_9$Cu(PO$_4$)$_6$O, arXiv:2308.00676 (2023).

\bibitem{ref6} H. Bai, L. Gao, J. Ye, C. Zeng, W. Liu, Magnetic Properties and Spin-orbit Coupling induced Semiconductivity in LK-99, arXiv:2308.05134 (2023).

\bibitem{ref7} K. Tao, R. Chen, L. Yang, J. Gao, D. Xue, C. Jia, The 1/4 occupied O atoms induced ultraflat band and the one dimensional channels in the Pb$_{10-x}$Cu$_x$(PO$_4$)$_6$O (x=0,0.5) crystal, arXiv:2308.03218 (2023).

\bibitem{ref8} Y. Sun, K.-M. Ho, and V. Antropov, Metallization and Spin Fluctuations in Cu-doped Lead Apatite, arXiv:2308.03454 (2023).

\bibitem{ref9} J. Hlinka, Possible ferroic properties of copper-substituted lead phosphate apatite, arXiv:2308.03691 (2023).

\bibitem{ref10} N. Mao, N. Peshcherenko, and Y. Zhang, Wannier functions, minimal model and charge transfer in Pb$_9$CuP$_6$O$_{25}$, arXiv:2308.05528 (2023).

\bibitem{ref11} M. Fidrysiak, A.P. Kądzielawa, and J. Spałek, High Temperature Superconductivity with Strong Correlations and Disorder: Possible Relevance to Cu-doped Apatite, arXiv:2308.03948 (2023).

\bibitem{ref12} D. M. Korotin, D. Y. Novoselov, A. O. Shorikov, V. I. Anisimov, and A. R. Oganov, Electronic correlations in promising room-temperature superconductor Pb$_9$Cu(PO$_4$)$_6$O: a DFT+DMFT study, arXiv:2308.04301 (2023).

\bibitem{ref13} S. V. Krivovichev, The crystal structure of Pb$_{10}$(PO$_4$)$_6$O revisited: the evidence of superstructure arXiv:2308.04915 (2023).

\bibitem{ref14} R. Kurleto, S. Lany, D. Pashov, S. Acharya, M. van Schilfgaarde, and D. S. Dessau, Pb-apatite framework as a generator of novel flat-band CuO based physics, including possible room temperature superconductivity, arXiv:2308.00698 (2023).

\bibitem{ref15} P. A. Lee and Z. Dai, Effective model for Pb$_9$Cu(PO$_4$)$_6$O, arXiv:2308.04480 (2023).

\bibitem{ref16} N. Chen, Y. Liu, and Y. Li, Statistical Material Law Support for Room Temperature Superconductivity in the Lead Apatite System, arXiv:2308.06349 (2023).

\bibitem{ref17} I. Panas, Entertaining the Possibility of RT Superconductivity in LK-99, arXiv:2308.06684 (2023).

\bibitem{ref18} H. Wu, L. Yang, B. Xiao, H. Chang, Successful growth and room temperature ambient-pressure magnetic levitation of LK-99, arXiv:2308.01516 (2023).

\bibitem{ref19} K. Kumar, N.K. Karn, and V.P.S. Awana, Synthesis of possible room temperature superconductor LK-99:Pb$_9$Cu(PO$_4$)$_6$O, arXiv:2307.16402 (2023).

\bibitem{ref20} Yi Jiang, et al., Pb$_9$Cu(PO$_4$)$_6$(OH)$_2$: Phonon bands, Localized Flat Band Magnetism, Models, and Chemical Analysis, arXiv:2308.05143 (2023).

\bibitem{ref21} Qiang Hou, Wei Wei, Xin Zhou, Yue Sun, Zhixiang Shi, Observation of zero resistance above 100 K in Pb$_{10-x}$Cu$_x$(PO$_4$)$_6$O, arXiv:2308.01192 (2023).

\bibitem{ref22} Qiang Hou, Wei Wei, Xin Zhou, Xinyue Wang, Yue Sun, ZhiXiang Shi, Current percolation model for the special resistivity behavior observed in Cu-doped Apatite, arXiv:2308.05778 (2023).

\bibitem{ref23} Kaizhen Guo, Yuan Li, and Shuang Jia, Ferromagnetic half levitation of LK-99-like synthetic samples, arXiv:2308.03110 (2023).

\bibitem{ref24} K. J. Prashant, Superionic phase transition of copper(I) sulfide and its implication for purported superconductivity of LK-99, arXiv:2308.05222 (2023).

\bibitem{ref25} P. Puphal, et al., Single crystal synthesis, structure, and magnetism of Pb$_{10-x}$Cu$_x$(PO$_4$)$_6$O, arXiv:2308.06256 (2023).

\bibitem{ref26} L. Liu et al., Semiconducting transport in Pb$_9$Cu(PO$_4$)$_6$O sintered from Pb$_2$SO$_5$ and Cu$_3$P, arXiv:2307.16802 (2023).

\bibitem{ref27} K. Kumar, N. K. Karn, and V. P. S. Awana, Absence of superconductivity in LK-99 at ambient conditions, arXiv:2308.03544 (2023).

\bibitem{ref28} S. Zhu, W. Wu, Z. Li, and J. Luo, First order transition in Pb9-xCux(PO4)O6 (0.9<x<1.1) containing Cu$_2$S, arXiv:2308.04353 (2023).

\bibitem{ref29} I. Timokhin et al., Synthesis and characterization of LK-99, arXiv:2308.03823 (2023).

\bibitem{ref30} G. S. Thakur, M. Schulze, and M. Ruck, On the synthesis methodologies to prepare Pb$_9$Cu(PO$_4$)$_6$O-phase, composition, magnetic analysis and absence of superconductivity, arXiv:2308.05776 (2023).

\bibitem{ref31} H. Singh et al., On the Experimental Evidence for Possible Room Temperature Superconductivity in LK-99, arXiv:2308.06589 (2023).

\bibitem{ref32} Y. Zhang, C. Liu, X. Zhu, and H.-H. Wen, Ferromagnetism and insulating behavior with a logarithmic temperature dependence of resistivity in Pb$_{10-x}$Cu$_x$(PO$_4$)$_6$O, arXiv:2308.05786 (2023).

\bibitem{ref33} L. Si and K. Held, Electronic structure of the putative room-temperature superconductor Pb$_9$Cu(PO$_4$)$_6$O, Phys. Rev. B 108, L121110 (2023).

\bibitem{ref34} L. Celiberti, L. Varrassi, and C. Franchini, Pb$_9$Cu(PO$_4$)$_6$O is a charge-transfer semiconductor, Phys. Rev. B 108, L201117 (2023).

\bibitem{ref35} D. M. Korotin, D. Y. Novoselov, A. O. Shorikov, V. I. Anisimov, and A. R. Oganov, Electronic correlations in the ultranarrow energy band compound Pb$_9$Cu(PO$_4$)$_6$O: A DFT+DMFT study, Phys. Rev. B 108, L241111 (2023).

\bibitem{ref36} J. Liu, T. Yu, J. Li, J. Wang, J. Lai, Y. Sun, X.-Q. Chen, and P. Liu, Symmetry breaking induced insulating electronic state in Pb$_9$Cu(PO$_4$)$_6$O, Phys. Rev. B 108, L161101 (2023).

\bibitem{ref37} O. Tavakol and T. Scaffidi, Minimal model for the flat bands in copper-substituted lead phosphate apatite: Strong diamagnetism from multiorbital physics, Phys. Rev. B 109, L100504 (2024).

\bibitem{ref38} J. Cabezas-Escareset al., Electronic structure and vibrational stability of copper-substituted lead apatite (LK-99), Phys. Rev. B 109, 144515 (2024).

\bibitem{ref39} Y. Sun et al., Metallization and spin fluctuations in Cu-doped lead apatite, Phys. Rev. Materials 7, 114804 (2023).

\bibitem{ref40} M.-L. Wang et al., Impact of sulfur doping on copper-substituted lead apatite, Phys. Rev. B 110, 104109 (2024).

\bibitem{ref41} S. M. Griffin, Origin of correlated isolated flat bands in copper-substituted lead phosphate apatite, arXiv:2307.16892 (2023).

\bibitem{ref42} J. Lai, J. Li, P. Liu, Y. Sun, and X.-Q. Chen, First-principles study on the electronic structure of Pb$_{10-x}$Cu$_x$(PO$_4$)$_6$ (x = 0, 1), arXiv:2307.16040 (2023).

\bibitem{ref43} D. Pashov et al., Multiple Slater determinants and strong spin-fluctuations as drivers of the insulating state in LK-99 within QSGW and beyond, arXiv:2308.09900 (2023).

\bibitem{ref44} L. Si, M. Wallerberger, A. Smolyanyuk, S. di Cataldo, J. Tomczak, and K. Held, Pb$_{10-x}$Cu$_x$(PO$_4$)$_6$O: a Mott or charge transfer insulator in need of further doping for (super)conductivity, J. Phys.: Condens. Matter 36, 065601 (2023).

\bibitem{ref45} L. Celiberti, L. Varrassi, and C. Franchini, Pb$_9$Cu(PO$_4$)$_6$ is a charge-transfer semiconductor, arXiv:2310.00006 (2023).

\bibitem{ref46} A. B. Georgescu, Why charge added using transition metals to some insulators, including LK-99, localizes and does not yield a metal, arXiv:2308.07295 (2023).

\bibitem{ref47} P. Puphal, M. Y. P. Akbar, M. Hepting, E. Goering, M. Isobe, A. A. Nugroho, and B. Keimer, Single crystal synthesis, structure, and magnetism of Pb$_{10-x}$Cu$_x$(PO$_4$)$_6$, APL Mater. 11, 101128 (2023).

\bibitem{ref48} K. Kumar, N. K. Karn, and V. P. S. Awana, Absence of Superconductivity in LK-99 at Ambient Conditions, ACS Omega 8, 43001 (2023).

\bibitem{ref49} K. Guo, Y. Li, and S. Jia, Ferromagnetic half levitation of LK-99-like synthetic samples, arXiv:2308.03110 (2023).

\bibitem{ref50} P. Wang et al., Ferromagnetic and insulating behavior in both half magnetic levitation and non-levitation LK-99 like samples, arXiv:2308.11768 (2023).

\bibitem{ref51} C. Liu et al., Phases and magnetism at microscale in compounds containing nominal Pb$_{10-x}$Cu$_x$(PO$_4$)$_6$O, Phys. Rev. Mater.7, 084804 (2023).

\bibitem{ref52} M.-L. Wang et al., Unveiling the impact of sulfur doping on copper-substituted lead apatite: A theoretical study, arXiv:2405.11854 (2024).

\bibitem{ref53} B. Cho, J. Park, and D. Yun, Exploration of superconductivity in LK-99 synthesized under different cooling conditions, Curr. Appl. Phys. 62, 22-28 (2024).

\bibitem{ref54} H. Wang et al., Indications of superconductivities in blend of variant apatite and covellite, arXiv:2406.17525v1 (2024)

\bibitem{ref55} H. Wang et al., Physical Zoo in Pb-Cu-P-S-O Apatite, Materials 18, 4728 (2025).

\bibitem{ref56} P. C. Hohenberg and B. I. Halperin, Theory of Dynamic Critical Phenomena, Rev. Mod. Phys. 49, 435 (1977).

\bibitem{ref57} K. Binder and A. P. Young, Rev. Spin glasses: Experimental facts, theoretical concepts, and open questions, Mod. Phys. 58, 801 (1986).

\bibitem{ref58} J. A. Mydosh, Spin glasses: redux: an updated experimental/materials survey, Rep. Prog. Phys. 78, 052501 (2015).

\bibitem{ref59} C. A. M. Mulder, A. J. van Duyneveldt, and J. A. Mydosh, Frequency and field dependence of the ac susceptibility of the CuMn spin-glass, Phys. Rev. B 25, 515 (1982).

\bibitem{ref60} J. Hammann, E. Vincent, V. Dupuis, M. Alba, M. Ocio, and J.-P. Bouchaud, Comparative review of aging properties in spin glasses and other disordered materials, J. Phys. Soc. Jpn. 69, Suppl. A, 206 (2000).

\bibitem{ref61} J. Souletie and J. L. Tholence, Critical slowing down in spin glasses and other glasses: Fulcher versus power law, Phys. Rev. B 32, 516 (1985).

\bibitem{ref62} Powders obtained after both synthesis stages have been characterized by x-ray diffraction (XRD) using a Rigaku SmartLab diffractometer. The XRD data were correlated with the energy-dispersive x-ray spectroscopy (EDS) data obtained using an Oxford Instruments X-Max 40 detector mounted on a Hitachi SU3500 scanning electron microscope.

\bibitem{ref63} B. Raveau, T. Sarkar, Superconducting-like behavior of the layered Chalcogenides CuS and CuSe below 40 K, Solid State Sci. 13, 1874 (2011).

\bibitem{ref64} I. Mazin, Structural and electronic properties of the two-dimensional superconductor CuS with 4/3-valent copper, Phys. Rev. B 85, 115133 (2012).

\bibitem{ref65} B. Raveau, T. Sarkar, and A. Pautrat, Comment on our article ``Superconducting-like behavior of the layered Chalcogenides CuS and CuSe below 40 K'', Solid State Sci. 14, 291 (2012).

\bibitem{ref66} V. Grinenko, A. Dudka, S. Nozaki, A. Muto, J. Clarke, J. Kilcrease, T. Hogan, V. Nikoghosyan, D. Napoletani, I. de Paiva, R. Dulal, S. Teknowijoyo, S. Chahid, and A. Gulian, Wohlleben effect and quantum criticality in YBa$_{1.4}$Sr$_{0.6}$Cu$_3$O$_6$Se$_{0.5}$ superconductor within a heterophase ceramic: Results for YBCO double-element substitution, Mod. Phys. Lett. B 39,  2540001 (2025)

\bibitem{ref67} Operationally, this strategy can be implemented by normalizing around a Cu-enriched baseline and scanning a ternary compositional neighborhood of the form Cu1+$\delta$-xMoxSy with $\delta>0$, $x\in[0,0.5]$, and $y$ remaining close to unity. The underlying hypothesis is that Mo incorporation introduces an additional degree of freedom to accommodate off-stoichiometric bonding constraints, effectively lowering decomposition energies of otherwise marginal compositions while preserving the high-Tc electronic characteristics associated with off-stoichiometric Cu$_x$S$_y$ phases.

\bibitem{ref68} P. Korbel and M. Novak, The Complete Encyclopedia of Minerals (Rebo International b.v. Lisse, the Netherlands) (1999).

\bibitem{ref69} For example, our sample in Fig. 1 was prepared via solid-state synthesis route.
\end{thebibliography}
\end{document}